\documentclass[%
 reprint,
 amsmath,amssymb,
 aps,
]{revtex4-2}

\usepackage{graphicx}% Include figure files
\usepackage{dcolumn}% Align table columns on decimal point
\usepackage{bm}% bold math

\usepackage{xcolor}
\usepackage{amsmath}
\usepackage[autostyle]{csquotes}

\usepackage{hyperref}
\hypersetup{colorlinks=true, linkcolor=blue, citecolor=blue, urlcolor=blue, unicode=false}

\begin{document}

\preprint{APS/123-QED}

\title{Contrasting electron-phonon interaction between electron- and hole-doped cuprates}

\author{Qinda Guo$^{1\star}$, Ke-Jun Xu$^{2,3,4\star}$, Magnus H. Berntsen$^{1}$, Antonija Grubi\v si\'c-\v Cabo$^{1,5}$, Maciej Dendzik$^{1}$, Thiagarajan Balasubramanian$^{6}$, Craig Polley$^{6}$, Su-Di Chen$^{2,3,4,7,8}$, Junfeng He$^{2,3,9}$, Yu He$^{2,3,10}$, Costel R. Rotundu$^{2,3}$, Young S. Lee$^{2,3,4}$, Makoto Hashimoto$^{11}$, Dong-Hui Lu$^{11}$, Thomas P. Devereaux$^{2,3,12\dagger}$, Dung-Hai Lee$^{13,14\dagger}$, Zhi-Xun Shen$^{2,3,4,15\dagger}$, and Oscar Tjernberg$^{1\dagger}$}

\affiliation{
\vspace{1mm}
\\$^{1}$Department of Applied Physics\mbox{,} KTH Royal Institute of Technology\mbox{,} Stockholm 11419\mbox{,} Sweden
\\$^{2}$Stanford Institute for Materials and Energy Sciences\mbox{,} SLAC National Accelerator Laboratory\mbox{,} 2575 Sand Hill Road\mbox{,} Menlo Park\mbox{,} California 94025\mbox{,} USA
\\$^{3}$Geballe Laboratory for Advanced Materials\mbox{,} Stanford University\mbox{,} Stanford\mbox{,} California 94305\mbox{,} USA
\\$^{4}$Department of Applied Physics\mbox{,} Stanford University\mbox{,} Stanford\mbox{,} CA 94305\mbox{,} USA
\\$^{5}$Zernike Institute for Advanced Materials\mbox{,} University of Groningen\mbox{,} 9747 AG Groningen\mbox{,} The Netherlands
\\$^{6}$MAX IV Laboratory\mbox{,} Lund University\mbox{,} Lund 22100\mbox{,} Sweden
\\$^{7}$Department of Physics\mbox{,} University of California\mbox{,} Berkeley\mbox{,} California 94720\mbox{,} USA
\\$^{8}$Kavli Energy NanoScience Institute\mbox{,} University of California\mbox{,} Berkeley\mbox{,} CA 94720\mbox{,} USA
\\$^{9}$Department of Physics and CAS Key Laboratory of Strongly-coupled Quantum Matter Physics\mbox{,} University of Science and Technology of China\mbox{,} Hefei\mbox{,} Anhui 230026\mbox{,} China
\\$^{10}$Department of Applied Physics\mbox{,} Yale University\mbox{,} New Haven\mbox{,} Connecticut 06511\mbox{,} USA
\\$^{11}$Stanford Synchrotron Radiation Lightsource\mbox{,} SLAC National Accelerator Laboratory\mbox{,} 2575 Sand Hill Road\mbox{,} Menlo Park\mbox{,} California 94025\mbox{,} USA 
\\$^{12}$ Department of Materials Science and Engineering\mbox{,} Stanford University\mbox{,} Stanford\mbox{,} California 94305\mbox{,} USA
\\$^{13}$Department of Physics\mbox{,} University of California\mbox{,} Berkeley\mbox{,} California 94720\mbox{,} USA
\\$^{14}$Material Sciences Division\mbox{,} Lawrence Berkeley National Laboratory\mbox{,} Berkeley\mbox{,} California 94720\mbox{,} USA
\\$^{15}$Department of Physics\mbox{,} Stanford University\mbox{,} Stanford\mbox{,} CA 94305\mbox{,} USA
\\$^{\star}$These authors contributed equally to this work.
\\$^{\dagger}$Corresponding authors: T.D. (\href{mailto:tpd@stanford.edu}{tpd@stanford.edu})\mbox{,} D.-H.L. (\href{mailto:dunghai@berkley.edu}{dunghai@berkley.edu})\mbox{,} Z.-X.S. (\href{mailto:zxshen@stanford.edu}{zxshen@stanford.edu})\mbox{,} O.T. (\href{mailto:oscar@kth.se}{oscar@kth.se})
}

%\date{\today}
\date{July 9, 2024}
\begin{abstract}
Spin- and charge-lattice interactions are potential key factors in the microscopic mechanism of high-temperature superconductivity in cuprates. Although both interactions can dramatically shape the low-energy electronic structure, their phenomenological roles in superconductivity are usually investigated independently. Employing angle-resolved photoemission spectroscopy, we reveal the spectroscopic fingerprint of short-range antiferromagnetic order in conjunction with enhanced electron-phonon interaction in the electron-doped cuprate superconductor $\mathrm{Nd_{1.85}Ce_{0.15}CuO_4}$. The observed mode coupling exhibits a strong momentum dependence that is in striking contrast to the node-antinode dichotomy previously observed in the hole-doped cuprates. Our results reveal an intimate relationship between electron-phonon coupling and antiferromagnetic fluctuations, which collectively sets the stage for unconventional superconductivity in the electron-doped cuprates.
\end{abstract}

\maketitle

Superconductivity in copper-oxide-based superconductors remains a topic of intense study, yet the origin and mechanisms underlying the high critical temperature ($T_\mathrm{c}$) remain elusive~\cite{bednorz1986possible}. Unlike conventional phonon-mediated superconductivity, as described by BCS theory~\cite{bardeen1957theory}, cuprate superconductors do not fit this picture, and electron-electron interaction-based pairing mechanisms are often considered as candidate coupling mechanisms~\cite{monthoux2007superconductivity,wang2011electron,scalapino2012common,o2022electron}. Electron-phonon interactions are, however, prevalent in cuprates~\cite{lanzara2001evidence,cuk2004coupling,tallon2005isotope,braden2005dispersion,park2008angle,anzai2017new} and have been shown to correlate with an amplified pairing strength in hole-doped cuprates~\cite{cuk2004coupling,devereaux2004anisotropic,he2018rapid}. Understanding the specific roles of and synergistic effects between the various interactions therefore seems essential for comprehending cuprate superconductivity.

In this context, the electron-doped cuprates serve as an excellent platform to study the interplay between magnetism and lattice interactions. Across a broad doping range, electron-doped cuprates exhibit robust antiferromagnetism (AF), with the superconducting dome emerging atop short-range AF order~\cite{armitage2010progress,sobota2021angle}. Moreover, unlike the hole-doped side, apart from antiferromagnetism there are no apparent intertwined or competing orders that may complicate interpretation. Nevertheless, although the existence of AF-related properties in electron-doped cuprates has been extensively studied and reported on, using multiple probing techniques~\cite{armitage2002doping,blumberg2002nonmonotonic,armitage2003angle,mang2004spin,braden2005dispersion,motoyama2007spin,he2019fermi}, the role of phonon interactions and the interplay with AF fluctuations have yet to be fully explored.

In this Letter, we present angle-resolved photoemission spectroscopy (ARPES) results on the low-energy electronic structure of the electron-doped cuprate superconductor $\mathrm{Nd_{2-n}Ce_{n}CuO_4}$ (NCCO) with \textit{n}=0.15. In the nodal direction, $(0,0)-(\pi,\pi)$, we identify a peak-dip-hump structure reminescent of what is commonly observed close to the antinode, $(\pi,0)$, in hole-doped cuprates. Near the antinode, a kink in the main dispersion with an energy scale of approximately 50~meV is observed that on the other hand resembles the nodal behavior in hole-doped cuprates. Both these phenomena are, in the case of cuprates, commonly tied to electron-phonon coupling~\cite{lanzara2001evidence,devereaux2004anisotropic,cuk2004coupling,shen2004missing,mishchenko2004electron,mannella2005nodal,lee2014interfacial}. Furthermore, both the peak-dip-hump structure and the kink strength show intimate interdependence with the manifested AF correlation strength as observed by the opening of an AF pseudogap and band folding across the AF zone boundary. In comparison with the hole-doped case ($\mathrm{Bi_{1.7}Pb_{0.4}Sr_{1.7}CaCu_{2}O_{8+\delta}}$ or PbBi-2212 at \textit{p}=0.186, $T_\mathrm{c}$=86~K), the spectroscopic fingerprint of the electron-phonon coupling exhibits a reversed dichotomy between node and antinode, yet a striking similarity in terms of the temperature dependence. The observed contrasting phenomenology in electron- and hole-doped materials gives new insights into the potential synergistic effects of magnetic correlations and electron-phonon coupling in cuprates.

Single crystals of NCCO were grown using the traveling solvent floating zone method with CuO flux in the molten zone. The as-grown NCCO crystals were annealed at 900~$^\circ$C under flowing Ar gas to optimize the superconducting properties, resulting in a $T_\mathrm{c}$ of $\sim 26$~K and a $\Delta T_\mathrm{c}$ of $\sim$2~K. Single crystals of PbBi-2212 were grown using the floating zone method with no additional flux, and the oxygen content was set by annealing in flowing N$_2$ at a temperature of 550 $^\circ$C. The $T_\mathrm{c}$ of PbBi-2212 is approximately 86~K, corresponding to a doping of $p\sim 0.186$. The $T_\mathrm{c}$ of the samples was determined by measuring AC susceptibility in a physical property measurement system, with the onset of diamagnetism taken as $T_\mathrm{c}$. The chemical composition of the crystals was determined by wavelength dispersive spectroscopy using an electron probe microanalyzer. The ARPES measurements were performed at the Bloch beamline (MAX IV Laboratory) for NCCO, and beamline 5-4 of the Stanford Synchrotron Radiation Lightsource for PbBi-2212. Single crystal samples were cleaved in situ using the top post method. Photon energies of 16.5~eV and 21~eV were employed for NCCO and PbBi-2212, respectively. The $s$-polarization of the light was used for the data collection. The beam spot size is estimated to be around 30~$\mathrm{\mu m}$ by 18~$\mathrm{\mu m}$ at the Bloch beamline of the MAX IV for the NCCO measurements. The energy resolution was approximately 16~meV for the NCCO measurements and 10~meV for the PbBi-2212 measurements, as determined by Fermi-edge measurements on a freshly evaporated polycrystalline gold film.

\begin{figure}[tbp]
\begin{center}
\includegraphics[scale=1.0]{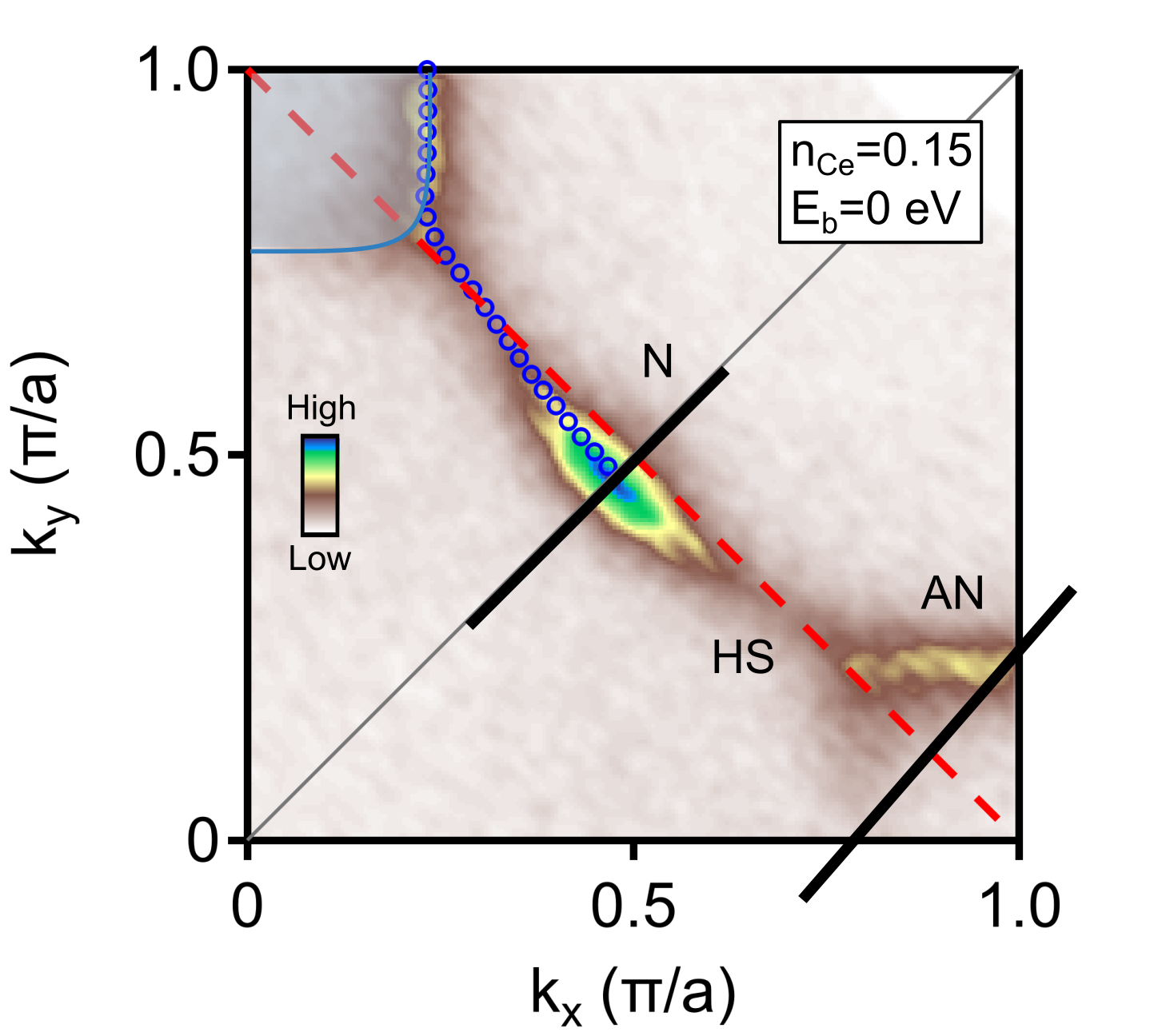}
\end{center}
\caption{Experimental constant energy surface (CES) map at $E_\mathrm{F}$ for $\mathrm{Nd_{1.85}Ce_{0.15}CuO_4}$. The data are symmetrized with respect to the zone diagonal (solid gray line) to cover a quarter of the first Brillouin zone. The black lines, denoted with N and AN show the momentum location of the energy-momentum cuts shown in Fig.~\textcolor{blue}{2}. HS denotes location of the hot spot. The blue circles indicate the fitted gossamer Fermi surface given by the intensity within the AF gap. The blue shaded area illustrate the electron-pocket arising from the AF correlations. The red dashed line indicates the AF zone boundary. The sample temperature was approximately 20~K.}
\label{fig:1}
\end{figure}

\begin{figure}[tbp]
\begin{center}
\includegraphics[scale=1.0]{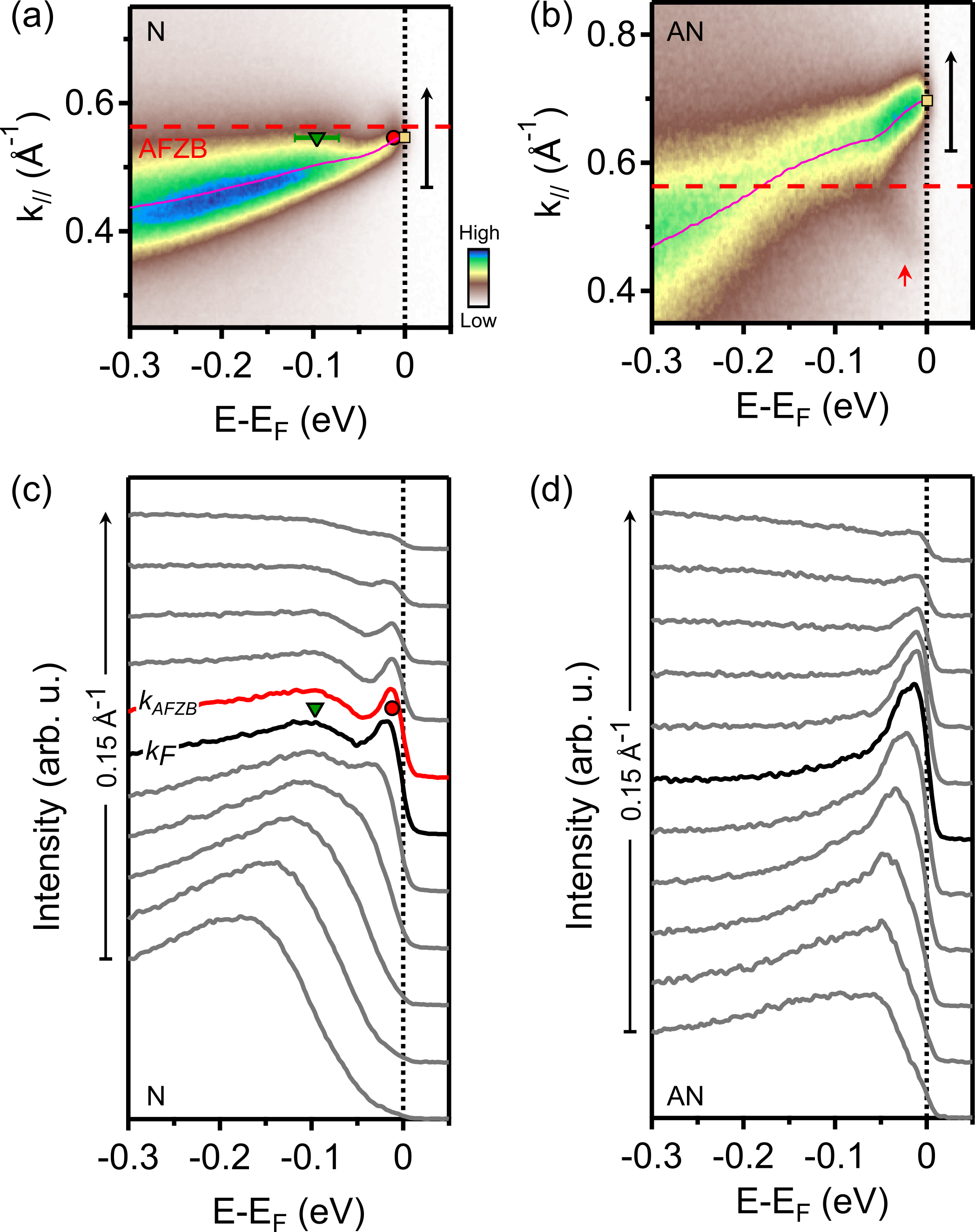}
\end{center}
\caption{Momentum-dependence of the low-energy spectra. Panels (a)-(b)  show energy-momentum cuts corresponding to cuts N and AN, respectively, as shown in Fig.~1. The solid purple lines represent the dispersion obtained from MDC fits. Their intersection with the Fermi energy determines the gossamer Fermi momentum. The red arrow in (b) indicates the folded band in relation to the AFZB. Panels (c)-(d) illustrate EDC stacks, with the range and direction indicated by black arrows in the corresponding upper panels, and the peak positions of the spectral features marked for $k_\mathrm{F}$-EDC in (c). The EDCs drawn with a thick black curve represents the momentum positions of the gossamer Fermi momentum and the red curve in (c) represents the momentum position of the AF zone boundary. The EDC curves are offset vertically for clarity. In (a) and (c), the red circles and green triangles mark the energy positions for the spectral features obtained from a multi-peak fit to the EDC at $k_\mathrm{F}$.}
\label{fig:2}
\end{figure}

The Fermi surface (FS) topology is fundamental for addressing various correlation effects. In the electron-doped cuprates, AF correlations induces an FS reconstruction with respect to the AF zone boundary (AFZB)~\cite{he2019fermi}. The finite magnitude of the AF order parameter leads to the opening of an AF pseudogap, and the fluctuation in the orientation of the AF order parameter gives rise to a low-energy spectral weight inside the pseudogap, leading to a \enquote{gossamer} FS that hosts superconducting coherence peaks in the superconducting state~\cite{xu2023bogoliubov}. In Fig.~\ref{fig:1}, a constant energy surface (CES) map at $E_\mathrm{b}$=0~eV ($\Delta E$=10~meV) is shown for a quarter of the first Brillouin zone for $\mathrm{Nd_{1.85}Ce_{0.15}CuO_4}$. The solid gray line indicates the nodal direction while the red dashed line delineates the AFZB. The spectral weight is substantially suppressed at the so called hot spot (HS) where the FS intersects with the AFZB. The blue circles in Fig.~\ref{fig:1} represent fitted peak positions of momentum distribution curves (MDC) at the Fermi energy and thus maps out the gossamer FS. Notably, the electron-pockets resulting from the AF reconstruction are evident, as outlined by the shaded area centered at (0,$\pi$) in the Brillouin zone. The thick black lines indicate momentum cuts near the nodal (N) and antinodal (AN) regions. The detailed band dispersion along these momentum cuts is depicted in Fig.~\ref{fig:2}.

The energy-momentum cuts at the node and antinode are presented in Fig.~\ref{fig:2}(a) and (b), respectively. For the nodal spectra, the ($\pi$,$\pi$) reconstruction of the electronic structure is discernible and appears as a back-folded feature symmetric around the AFZB near $E_\mathrm{F}$. Details of the AF dispersion are provided in the Supplementary Material (SM), section II. The momentum dependence as well as the energy scale of this feature shows consistency with previous reports~\cite{matsui2005angle,matsui2005direct,matsui2007evolution,park2013interaction,he2019fermi,xu2023bogoliubov}. A stack of energy distribution curves (EDCs) extracted from the momentum resolved spectra in panels (a) and (b) are presented in panels (c) and (d), respectively, enabling a closer examination and comparison of the spectral features along the nodal and anti-nodal cuts. The momentum range of the EDC stacks are indicated by the vertical black arrows in panels (a) and (b) respectively, and corresponds to a range of 0.15~$\mathrm{\AA}^{-1}$ centered around the gossamer Fermi momentum, $k_\mathrm{F}$, which is marked with a yellow square in panels (a) and (b). The spectra in the nodal direction display a clear peak-dip-hump structure in the vicinity of the AFZB. This peak-dip-hump structure is reminiscent of the peak-dip-hump structure observed at the anti-node in hole doped compounds but have previously not been observed in NCCO \cite{schmitt2008analysis,he2019fermi}. A likely explanation for the present observation is the improved experimental conditions, particularly the combination of a small beam spot, high momentum resolution, and high sample quality. A quantitative analysis of the line shape for $k_\mathrm{AFZB}$ was carried out by employing multi-Lorentzian fits to the EDC (see SM for further details). The determined peak and hump positions at the Fermi momentum are displayed as red and green symbols in panels (a) and (c). We note that the dip of the peak-dip-hump structure is located at approximately 50~meV binding energy. In contrast, the broad hump feature is absent in the EDCs from the antinodal cut as seen in panel (d). Instead, a kink at a binding energy of approximately 50~meV is evident in the main band in panel (b). The occurrence of a kink in the band dispersion near the antinode is consistent with prior findings in electron-doped cuprates~\cite{park2008angle, schmitt2008analysis, liu2012identification}. This type of kink in the dispersion is commonly interpreted as a signature of electron-phonon interaction. With only one kink resolved in the main dispersion, and the energy scale aligning well with that of the dip, it is plausible that the same phonon (or phonons) is responsible for the dispersion anomalies at both the node and antinode. While other explanations for the peak-dip-hump structure can be envisaged, such as coupling to spin excitations, the energy range of known spin excitations does not cover the 50~meV range \cite{zhao2007neutron,yu2010two}. The similar energy of the dip and the anti-nodal kink makes electron-phonon coupling a compelling explanation. Also visible at the antinode is the band-folding (marked with a red arrow in Fig.~\ref{fig:2}(b)) across the AFZB, which persists and is seen as an additional dispersing low energy feature in panel (b), leading to the formation of the electron-like pocket around the AFZB.

\begin{figure}[tbp]
\begin{center}
\includegraphics[scale=1.0]{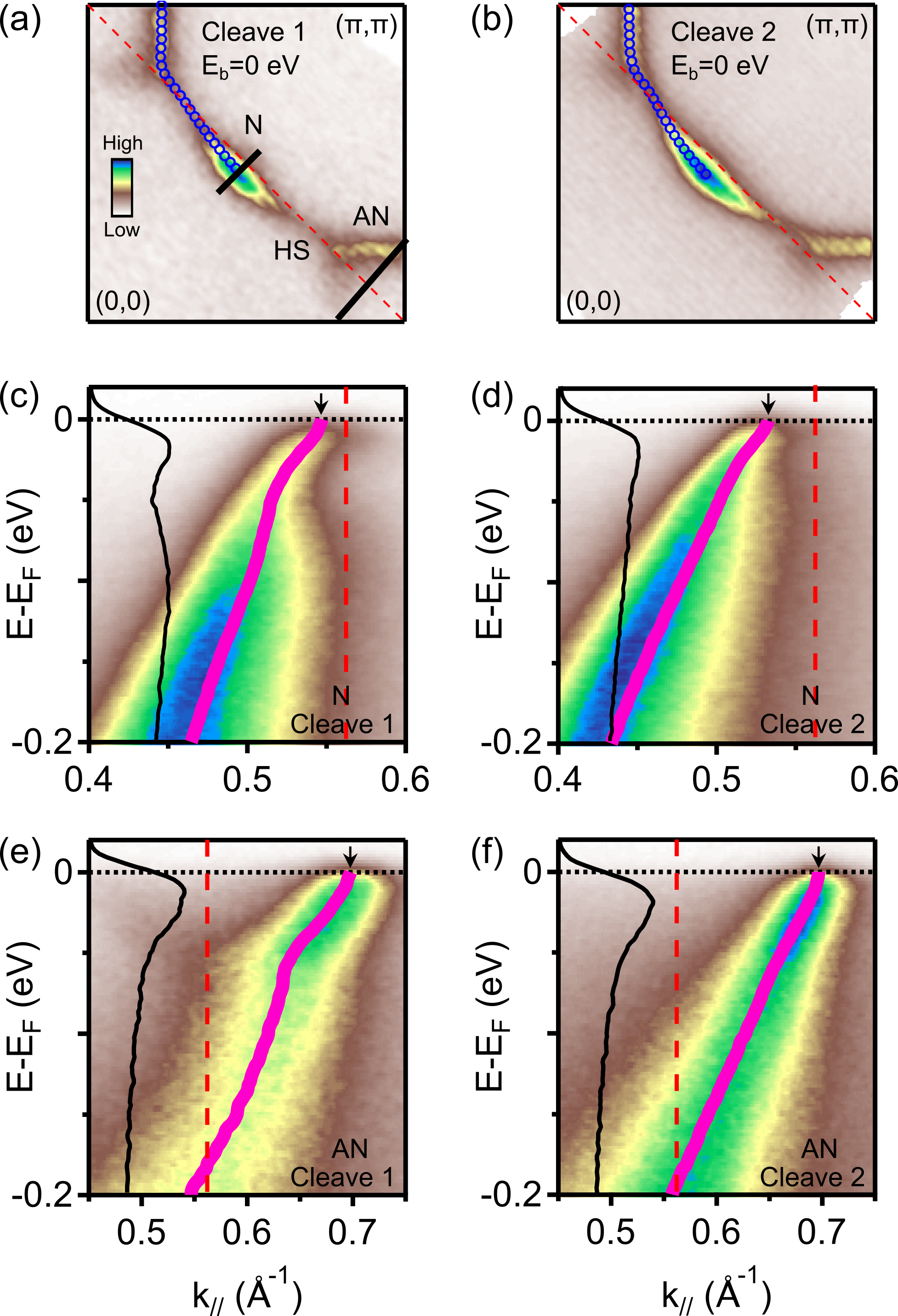}
\end{center}
\caption{The interplay between electron-phonon coupling and AF fluctuation. Panels (a) and (b) show the CES mapping at E$_\mathrm{F}$ for two different cleaves denoted 1 and 2. (c) and (d) illustrate the nodal energy-momentum cuts along the direction of N as indicated in panel (a) for cleaves 1 and 2, respectively. Similarly, (e) and (f) show the antinodal cuts along AN for the two cleaves. In each panel of (c)-(f), the $k_\mathrm{F}$-EDCs taken at the position marked by the black arrow are plotted as a black curve on the left edge of each panel. The sample temperature was approximately 20~K.}
\label{fig:3}
\end{figure}

To investigate the interplay between the electron-lattice coupling and AF, we studied the dependence of the electronic structure on the AF correlation strength. In the electron-doped cuprates, the AF properties are known to be very sensitive to the annealing conditions~\cite{tokura1989superconducting,horio2016suppression}.  The AF correlation strength can also vary substantially from cleave to cleave in samples with the same doping and even between successive cleaves of the same sample. An increase in the AF correlation strength can be observed as stronger band folding and increased suppression of the spectral weight at the HS. Figure~\ref{fig:3} compares data obtained from two different cleaves of the same bulk sample, denoted as cleave 1 and cleave 2. The data previously shown in Fig.~\ref{fig:1} and Fig.~\ref{fig:2} are obtained from cleave 1. Fermi surface maps for the two cleaves are displayed in Fig.~\ref{fig:3}(a) and (b), with the corresponding fitted gossamer FS overlaid as blue circles. The calculated area enclosed by the fitted Fermi surfaces yields nominal carrier densities of 15.6$\pm$0.4\% and 15.4$\pm$0.5\% for cleave 1 and 2, respectively. The band dispersion along the nodal and antinodal cuts (purples curves) for the two cleaves are presented in panels (c)-(f), with corresponding labels provided for clarity. Near the left edge of each intensity plot, the $k_\mathrm{F}$-EDCs are shown as black curves, extracted at momentum positions indicated by black arrows.

While the precise origin of the variations in AF correlation strength is not known, we stress that the above FS analysis shows no variation in doping to within the error bars of the FS fits. The AF correlation is notably stronger in cleave 1 compared to cleave 2, as evidenced by the following observations: i) In terms of the near-$E_\mathrm{F}$ dispersion at the node, the AF-folding is much more pronounced in cleave 1, corroborated by a stronger renormalization of the Fermi velocity near $E_\mathrm{F}$. ii) The near-$E_\mathrm{F}$ intensity suppression at the HS due to AF fluctuations is noticeably stronger in cleave 1 (see SM for further details). iii) The AF-folded branch at the antinode is more prominent in cleave 1. Concurrently, the peak-dip-hump structure at the node and the kink feature at the antinode become more pronounced from cleave 2 to cleave 1. This indicates an apparent increase in the electron-phonon interaction with increasing AF correlations. Lastly, the shape of the electron pocket appears to become more rectangular in the presence of stronger AF correlations, as evidenced in panels (a) and (b) in Fig.~\ref{fig:3}. While it is conceivable that the observed variation in AF signatures detailed above could result from differences in surface scattering connected to variations in surface quality, MDC analysis of near-$E_\mathrm{F}$ spectral features (see SM section III) shows greater momentum broadening for cleave 1 compared to cleave 2. This indicates that a somewhat more irregular surface promotes stronger AF correlations rather than diminishing them.

\begin{figure}[tbp]
\begin{center}
\includegraphics[scale=1.0]{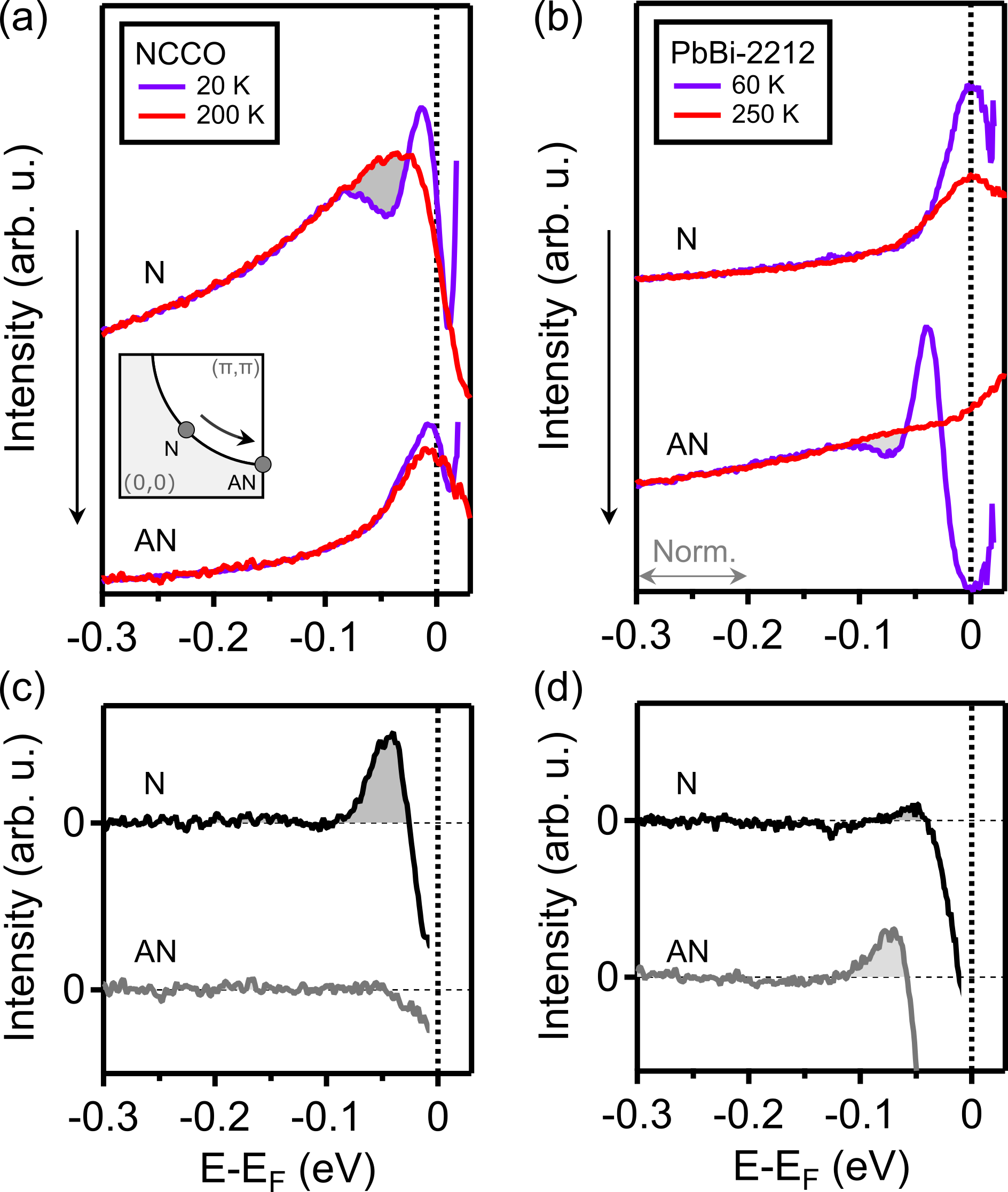}
\end{center}
\caption{Comparison between the electron- and hole-doped cuprates. (a) The $k_\mathrm{F}$-EDCs at nodal and antinodal directions are shown. The measurement temperatures are 20~K and 200~K, corresponding to the purple and red curves, respectively. The sample is $\mathrm{Nd_{1.85}Ce_{0.15}CuO_4}$ ($n=0.15$, $T_\mathrm{c}$=26~K). (b) Similar to (a), but with results from Bi$_{1.7}$Pb$_{0.4}$Sr$_{1.7}$CaCu$2$O$_{8+\delta}$ ($p=0.186$, $T_\mathrm{c}$=86~K). The measurement temperatures in this case are 60K and 250~K. (c) and (d) show the dip spectral weight obtained from the difference between corresponding $k_\mathrm{F}$-EDCs in (a) and (b).}
\label{fig:4}
\end{figure}

Spectroscopically, both the electron and hole-doped cuprates manifest the nodal-antinodal dichotomy. In the electron-doped cuprate NCCO, our data demonstrate a stark contrast between the observed peak-dip-hump structure at the node and the kink at the antinode. In the hole-doped cuprates, on the other hand, the opposite behavior is observed ~\cite{campuzano1999electronic,lanzara2001evidence,borisenko2003anomalous,cuk2004coupling,devereaux2004anisotropic,he2018rapid,chen2019incoherent}, where kinks are observed at the node, but a strong peak-dip-hump structure is observed at the antinode. It is worth noting that in the Bi2212 system, several phonons couple to the node, and the major dispersion kink (70-80~meV binding energy) is typically attributed to the in-plane oxygen breathing mode \cite{cuk2004coupling,devereaux2004anisotropic}. In Fig.~\ref{fig:4}, we compare the line shapes observed for NCCO with the ones observed on the hole-doped side, here represented by PbBi-2212. EDCs at the Fermi momentum from NCCO ($\mathrm{Nd_{2-n}Ce_{n}CuO_4}$, \textit{n}=0.15, $T_\mathrm{c}$=26~K) and PbBi-2212 (Bi$_2$Sr$_2$CaCu$2$O$_{8+\delta}$, \textit{p}=0.186, $T_\mathrm{c}$=86~K) are shown in panels (a) and (b), respectively. The results obtained near the node and antinode for both samples are arranged from top to bottom in each panel. Each EDC has been divided by the Fermi-Dirac distribution, and the intensity has been normalized within the energy window of -0.3 to -0.2~eV. The high binding energy part of the spectrum is the least effected by the spectral weight transfer around the Fermi level and therefore chosen for normalization \cite{chen2019incoherent}. The specific range chosen for normalization has little influence on the qualitative nature of the difference spectra. The data presented were measured at 20~K and 200~K for NCCO, and at 60~K and 250~K for PbBi-2212. As the temperature is lowered for NCCO, a redistribution of the spectral weight at the node occurs and the quasiparticle peak near the antinode in NCCO sharpens. In contrast, for PbBi-2212, the superconducting gap at the antinode reaches $~$40~meV at low temperatures, consistent with previous results \cite{hashimoto2014energy}, and the peak-dip-hump spectral shape emerges. This is further underscored by extracting the spectral weight of the \enquote{dip} (area highlighted in gray) by computing the difference between the high and low temperature $k_\mathrm{F}$-EDCs. The result is illustrated in panels (c) and (d) for NCCO and PbBi-2212, respectively. In Bi-2212, the combined anisotropic effect in the band structure and the d-wave gap results in a strong nodal and anti-nodal anisotropy when considering the peak-dip-hump structure for an isotropic Holstein model \cite{sandvik2004effect}. We have tried to reproduce the reversed dichotomy in NCCO using the same approach, where the AF gap in NCCO takes the same role as the superconducting gap in Bi-2212. However, we were unable to account for the reversed dichotomy in this way. A notable difference in the electron-phonon coupling between NCCO and PbBi-2212 is related to the differing crystal structures. While NCCO has the $T^{\prime}$ structure~\cite{takagi1989superconductivity,armitage2010progress}, which is characterized by the absence of apical oxygen, PbBi-2212 crystallizes in the $T$ structure with apical oxygen. As a consequence, the $A_\mathrm{1g}$ and $B_\mathrm{1g}$ phonons, while having the relevant energy scale in both compounds, represent different atomic motions.

The apparent change in electron-phonon coupling observed in panels (e) and (f) of Fig.~\ref{fig:3} points to a direct influence of the AF fluctuations on the electron-phonon interaction since there is no AF-induced gap close to $E_\mathrm{F}$ at the antinode and hence only a minor redistribution of spectral weight due to AF folding. Such an AF-related enhancement of the electron-phonon coupling has been indicated in previous works~\cite{lupi1998fano,lupi1999evolution,macridin2006synergistic,gunnarsson2008interplay,franchini2021polarons,cai2021antiferromagnetism}. Specifically, it has been demonstrated that a short-range AF background can promote the formation of lattice polarons in electron-doped cuprates~\cite{cappelluti2009polaronic}. This aligns with the observations presented here.

In the case of hole-doped cuprates, the spectral dip has been attributed to coupling to a $B_\mathrm{1g}$ phonon mode~\cite{scalapino1966strong,cuk2004coupling,he2018rapid}. The coupling is suggested to contribute to spectral weight depletion, as the phonon acquires some electronic character, and thus the spectral weight and the extent of depletion becomes a sign of the coupling strength. When interpreting the effect of DOS pile-up in PbBi-2212, the flat band dispersion at the antinode in this system and the proximity to the van Hove singularity need to be considered~\cite{he2021superconducting}. In NCCO, the approximate 50~meV energy scale of the dip feature suggests that the phonon (or phonons) involved are oxygen related. However, the specifics of the electron-phonon coupling require a detailed theoretical analysis, similar to the one conducted for Bi-2212~\cite{devereaux2004anisotropic}. While an enhanced coupling to the $B_\mathrm{1g}$ phonon in Bi-2212 has been shown to coincide with enhance d-wave superconductivity~\cite{he2018rapid}, the role of phonons in NCCO remains an open question. Nevertheless, it is clear that the amount of AF correlations and the related changes in the electronic structure, such as band folding and gap opening, is delicately balanced in NCCO. At the same time, there is an intimate link between the AF correlations and the electron-phonon coupling as shown in Fig.~\ref{fig:3}. It is therefore a tantalizing question whether electron-phonon coupling plays a stabilizing and enhancing role in the emergence of superconductivity in NCCO.

In summary, we have investigated the low-energy electronic structure in the electron-doped cuprate superconductor $\mathrm{Nd_{1.85}Ce_{0.15}CuO_4}$. The spectroscopic signatures of short-range AF order and electron-boson interaction are clearly resolved in momentum space. Significantly, we observe a distinct peak-dip-hump structure along the (0,0)-($\pi$,$\pi$) direction in this system. Conversely, we observe a kink in the band dispersion near the antinode. This node-antinode dichotomy reveals the anisotropic nature of the electron-phonon interaction. Furthermore, a close interplay between AF correlations and phonon coupling is observed, suggesting the presence of spin-lattice interactions. Comparing to data on the PbBi-2212 system, we demonstrate that the two systems display a similar peak-dip-hump structure but with the behavior at the node and antinode switched. Overall, our findings point to a considerable and momentum dependent electron-phonon coupling that is strongly coupled to spin correlations in cuprates. These observations raise questions about a possible synergistic effect between AF spin fluctuations and electron-phonon coupling for the emergence of cuprate superconductivity. While it seems likely that the underlying mechanism(s) behind superconductivity is common to all cuprates, it cannot be excluded that the details of the interplay between different orders and couplings are different for electron and hole doped cuprates.

We acknowledge the MAX IV Laboratory for beamtime on Bloch Beamline under Proposal 20220426. The work at KTH and MAX IV was supported by Swedish Research Council grants 2019-00701 and 2019-03486, as well as The Knut and Alice Wallenberg foundation grant 2018.0104. Use of the Stanford Synchrotron Radiation Lightsource, SLAC National Accelerator Laboratory, is supported by the U.S. Department of Energy, Office of Science, Office of Basic Energy Sciences under Contract No. DE-AC02-76SF00515. Q. Guo, A. Grubisic-Cabo, M. Dendzik acknowledge support by KTH Materials Platform. M. Dendzik acknowledges financial support from the Göran Gustafsson Foundation and the Swedish Research Council under Grant No. 2022-03813. D.-H.Lee. was supported by the US Department of Energy, Office of Science, Basic Energy Sciences, Materials Sciences and Engineering Division, contract no. DE-AC02-05-CH11231 within the Quantum Materials Program (KC2202).

\nocite{*}

\bibliography{ref}
\end{document}

\centerline{\large{Supplemental Material for}}

\title{Contrasting electron-phonon interaction between electron- and hole-doped cuprates}

\author{Qinda Guo$^{1\star}$, Ke-Jun Xu$^{2,3,4\star}$, Magnus H. Berntsen$^{1}$, Antonija Grubi\v si\'c-\v Cabo$^{1,5}$, Maciej Dendzik$^{1}$, Thiagarajan Balasubramanian$^{6}$, Craig Polley$^{6}$, Su-Di Chen$^{2,3,7,8}$, Junfeng He$^{2,3,4,9}$, Yu He$^{2,3,10}$, Costel R. Rotundu$^{2,3}$, Young S. Lee$^{2,3,4}$, Makoto Hashimoto$^{11}$, Dong-Hui Lu$^{11}$, Thomas P. Devereaux$^{2,3,12\dagger}$, Dung-Hai Lee$^{13,14\dagger}$, Zhi-Xun Shen$^{2,3,4,15\dagger}$, and Oscar Tjernberg$^{1\dagger}$}

\affiliation{
\vspace{1mm}
\\$^{\star}$These authors contributed equally to this work.
\\$^{\dagger}$Corresponding authors: T.D. (\href{mailto:tpd@stanford.edu}{tpd@stanford.edu})\mbox{,} D.-H.L. (\href{mailto:dunghai@berkley.edu}{dunghai@berkley.edu})\mbox{,} Z.-X.S. (\href{mailto:zxshen@stanford.edu}{zxshen@stanford.edu})\mbox{,} O.T. (\href{mailto:oscar@kth.se}{oscar@kth.se})
}

\pacs{74.72.Gh, 71.38.-k, 74.78.-w}

\maketitle

\newpage

%%%%%%%%%%%%%%%%%% Part 1 %%%%%%%%%%%%%%%%%%%%%%%%%%%%%%%%%%%%%%%%%%%%%%
{\bf I. Measurement of dynamic susceptibility}

\begin{figure*}[h]
\begin{center}
\includegraphics[scale=1.0]{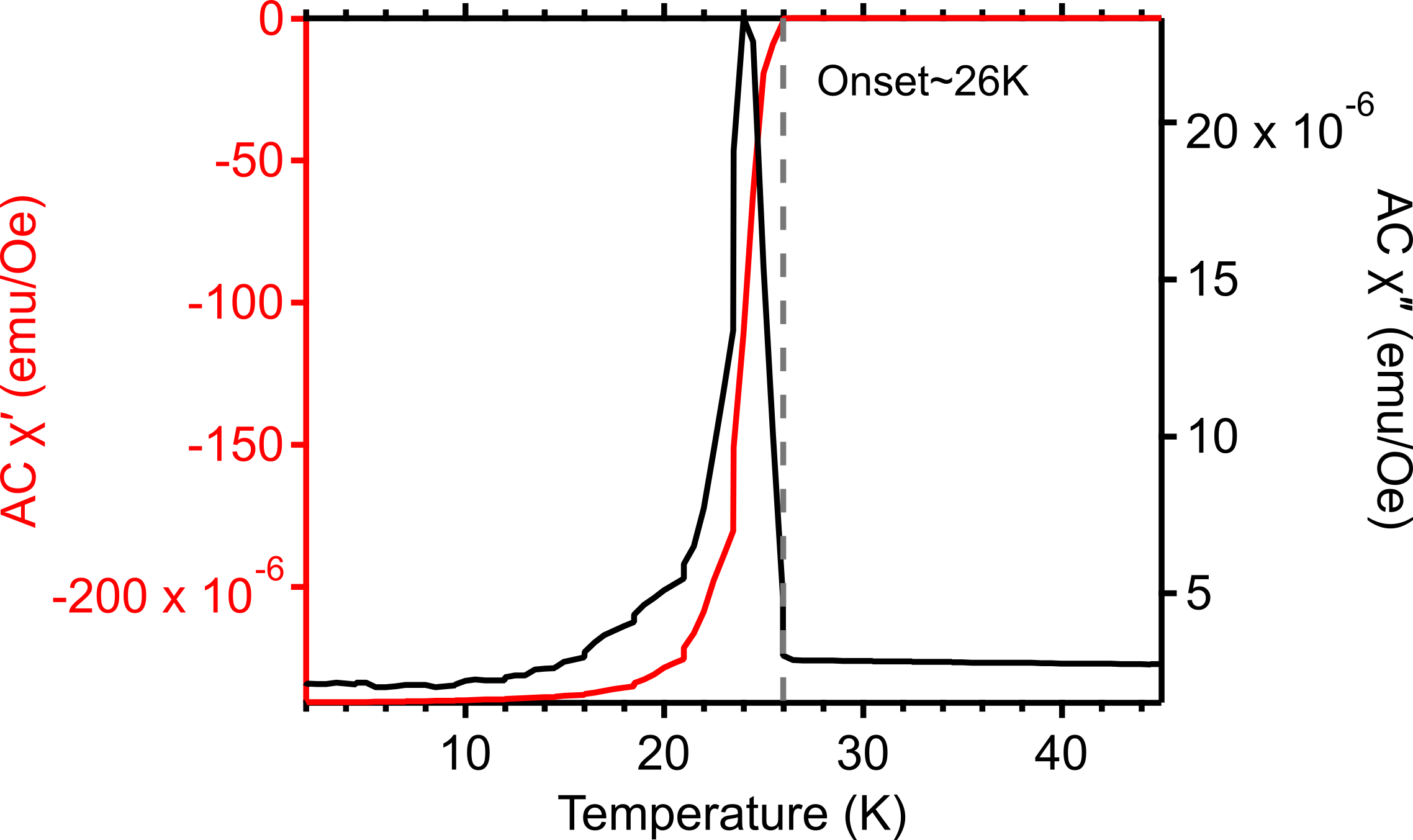}
\end{center}
\caption{\textbf{AC magnetic susceptibility measurements of the NCCO sample.} The magnetic field amplitude is 5~Oe. The real component of the susceptibility is depicted in red, whereas the imaginary part is shown in black.}
\label{figs:1}
\end{figure*}

The superconducting properties of the NCCO sample was characterized through a Physical Property Measurement System, using the AC magnetization option. In this work, we adhere to the so-called conventional annealing method~[1] and focus on the nominal doping level of 0.15, which yields the highest $T_\mathrm{c}$ in NCCO. Figure~\ref{figs:1} illustrates the temperature dependence of the real and imaginary components of the AC susceptibility, $\chi$, during zero-field cooling. The AC field amplitude was 5~Oe. The onset of diamagnetism indicated by $\chi '$ is taken as $T_\mathrm{c}$, whereas $\chi ''$ serves as a reference. Notably, the superconducting transition appears clean, occurring only at 26~K. This suggests that the sample bulk is homogeneous, which is also supported by the volume counting of the fitted Fermi surface as described in the main text. Therefore, the observed variation in magnetic correlation strength upon cleaving is unlikely due to significant bulk inhomogeneity. In this context, several studies have proposed a uniform state, ruling out mesoscopic phase separation, in the electron-doped cuprates~[2-4]. \\

%%%%%%%%%%%%%%%%%% Part 2 %%%%%%%%%%%%%%%%%%%%%%%%%%%%%%%%%%%%%%%%%%%%%%

{\bf II. The AF back-bending behavior along the zone diagonal}

\begin{figure*}[tbp]
\begin{center}
\includegraphics[scale=1.0]{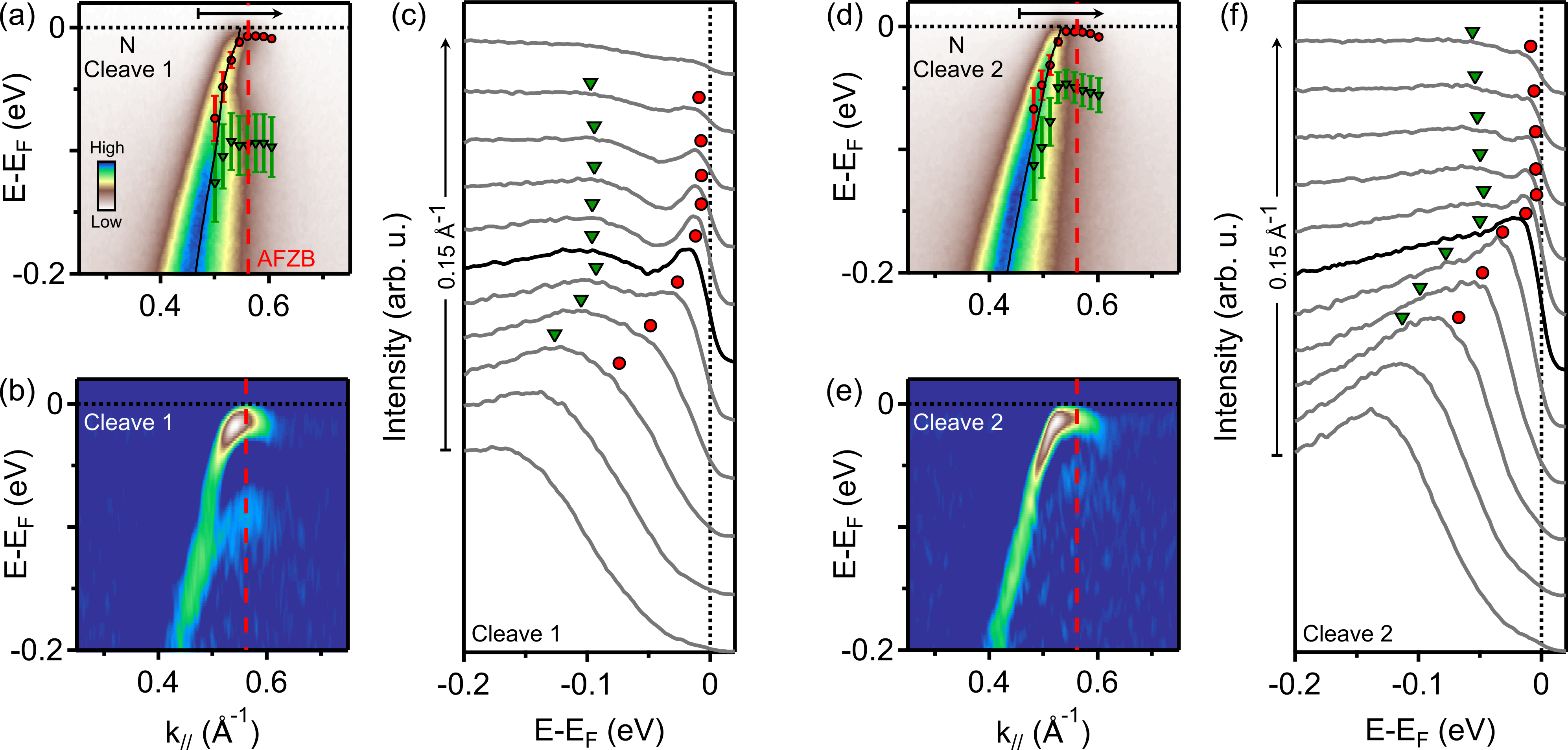}
\end{center}
\caption{\textbf{The electronic structure along the zone diagonal for the two different cleaves.} (a) The energy-momentum cuts in cleave 1, which are taken along (0,0)-($\pi$,$\pi$) direction. The red dashed line indicates the AF zone boundary. The red circles and green triangles mark the energy positions obtained from multipeak fits to the energy distribution curves (EDC) presented in (c). (b) Second derivative (with respect to energy) of the intensity plot in panel (a). (c) EDCs extracted from (a) with the peak positions of the spectral features marked. The black curve represents the momentum positions of the gossamer Fermi momentum. The range and direction of the EDC stack are marked by the black arrow on top of (a). The results from cleave 2 are demonstrated similarly in panels (d)-(f). The measurement temperature was 20~K.}
\label{figs:2}
\end{figure*}

\begin{figure*}[tbp]
\begin{center}
\includegraphics[scale=1.0]{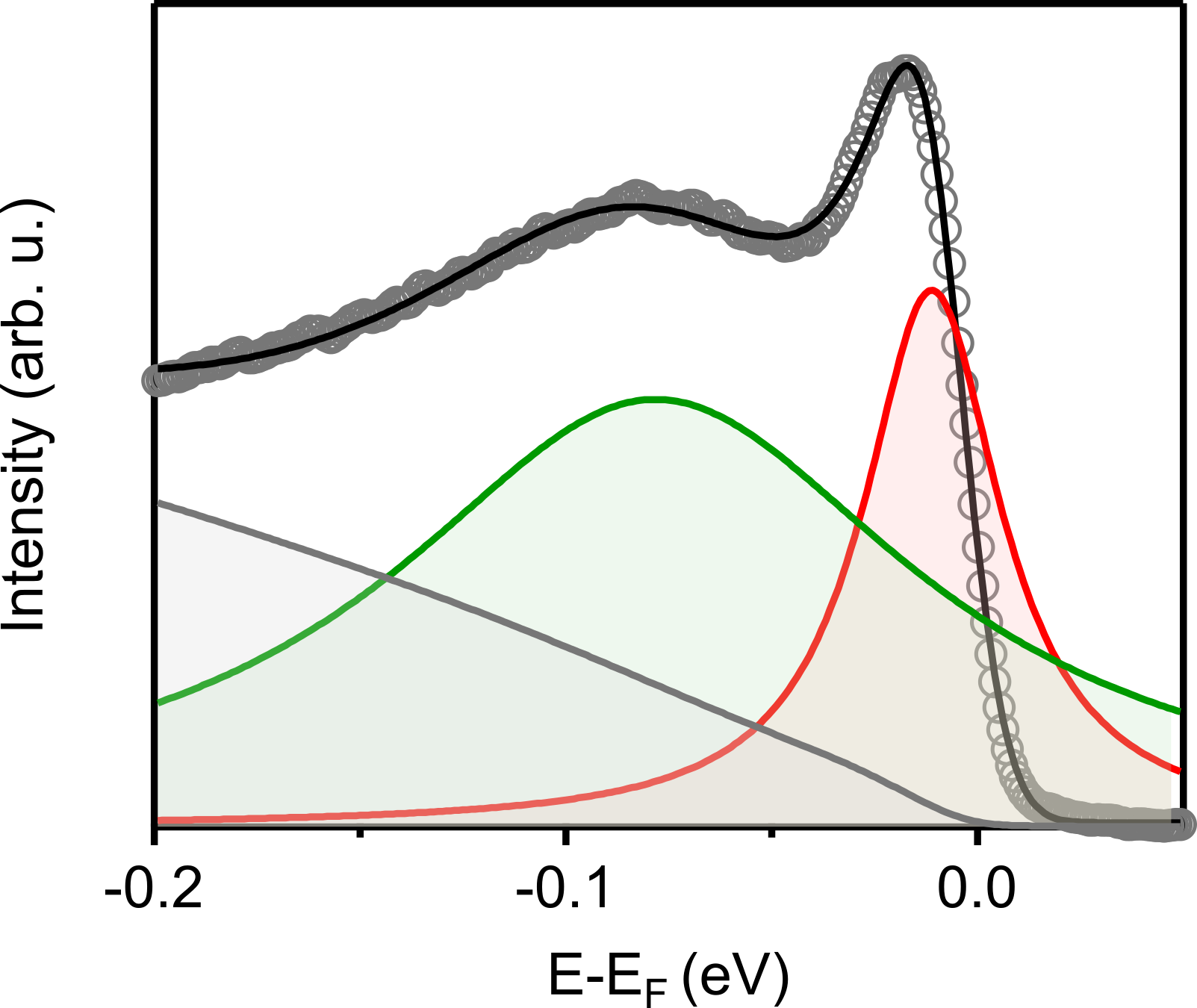}
\end{center}
\caption{\textbf{An example of the multi-peak fitting to the energy distribution curves (EDC).} The raw data is depicted using gray circles. The two features are denoted by the red and green Lorentzian peaks. The integrated background is shown as a solid gray line, while the sum of the fitted components multiplied by a Fermi function is illustrated by the black solid line.}
\label{figs:3}
\end{figure*}

In the main text, we demonstrate the Fermi surface (FS) reconstruction near the AF zone boundary, in terms of folded intensity and the visibility of electron pockets. In this section, we investigate the AF-folding along the zone diagonal in greater detail for the two cleaves in Fig.~\ref{figs:2}. The purpose is to showcase the AF feature, as marked by the red circles, exhibiting a clear back-bending near $E_\mathrm{F}$ with an energy scale of about 10~meV. This is corroborated by the second derivative plots as well as the energy distribution curve (EDC) stacks. We note that in both cleaves, the band folds at a similar energy scale. The dispersion of the broad hump feature is indicated by the green triangles. The details of the fitting method provided in Fig.~\ref{figs:3}. However, in cleave 2, which has a much weaker coupling strength, the energy separation between the peak and hump features is smaller and approximately equal to a single phonon energy. \\

%%%%%%%%%%%%%%%%%% Part 3 %%%%%%%%%%%%%%%%%%%%%%%%%%%%%%%%%%%%%%%%%%%%%%

{\bf III. Comparing the AF strength across different cleaves}

\begin{figure*}[tbp]
\begin{center}
\includegraphics[scale=1.0]{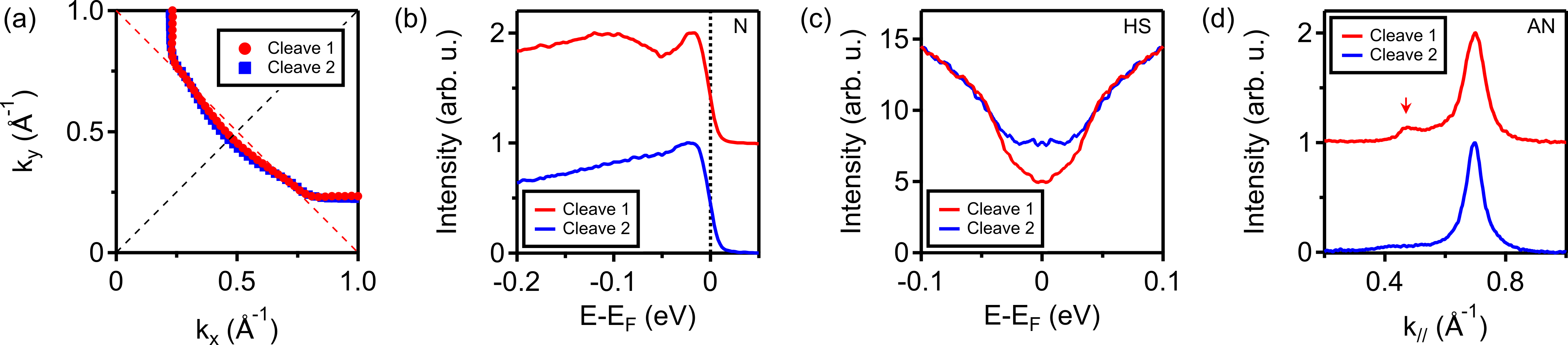}
\end{center}
\caption{\textbf{The comparison of the AF correlation between the two different cleaves.} (a) Fitted gossamer Fermi surface at $E_\mathrm{F}$ for cleaves 1 and 2, respectively. The data are symmetrized with respect to the zone diagonal (black dashed line) to cover a quarter of the first Brillouin zone. The red dashed line indicates the AF zone boundary. The extracted $k_\mathrm{F}$-EDCs at the node and the hot spot are shown in (b) and (c), respectively. In (c), the EDCs are symmetrized around the Fermi level. (d) The MDCs integrated at $E_\mathrm{b}$=0~eV ($\Delta E$=10~meV). The red arrow indicates the folded intensity attributed to the AF order.}
\label{figs:4}
\end{figure*}

In this section, we present experimental evidence indicating that cleave 1 exhibits a significantly higher AF order parameter amplitude compared to cleave 2. In Fig.~\ref{figs:4}(a)-(d), we show parallel comparisons between the cleaves, with a focus on the fitted gossamer FS, nodal $k_\mathrm{F}$-EDCs, hot spot $k_\mathrm{F}$-EDCs, and antinodal momentum distribution curves (MDCs) taken at the Fermi energy ($E_\mathrm{F}$). Specifically, in panel (a), one can observe that the two Fermi surfaces overlap well, despite subtle differences along the zone diagonal and near the zone boundary. The gossamer $k_\mathrm{F}$ is closer to AFZB in cleave 1 along the zone diagonal, and the electron pocket appears to be more rectangular near the zone boundary. In panel (b), the peak-dip-hump structure is more pronounced in cleave 1, however, the energy position of the dip is similar. Regarding the hot spot $k_\mathrm{F}$-EDCs depicted in (c), the in-gap intensity at $E_\mathrm{F}$ measures approximately 4.7 for cleave 1 and 7.7 for cleave 2, indicating a more pronounced suppression due to AF fluctuations. In panel (d), the MDC integrated at $E_\mathrm{F}$ suggests a noticeable folded intensity for cleave 1 compared to cleave 2. The intensity ratio (shadow band intensity / main band intensity) is approximately 13.7\% for cleave 1 and 6\% for cleave 2. 

\begin{figure*}[tbp]
\begin{center}
\includegraphics[scale=1.0]{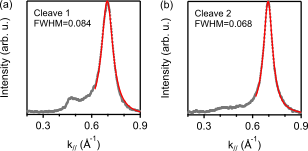}
\end{center}
\caption{\textbf{The MDC width analysis for the two cleaves.} The gray dots represent the MDC taken at the Fermi level, while the red curve shows the fitting result.}
\label{figs:5}
\end{figure*}

Additionally, the MDCs extracted at the Fermi level ($\mathrm{\Delta E}$ = $\pm$5 meV) for cleaves 1 and 2 are presented in Fig.~\ref{figs:5}. Apart from the clearer band folding in cleave 1, which is attributed to stronger AF correlation, the width of the main peak is slightly broader in cleave 1 compared to cleave 2. Given the surface-sensitive nature of ARPES, surface scattering, which can vary from cleave to cleave may dilute delicate spectroscopic features. Such a cleave dependent effect could in principle lead to the simultaneous diminishing of apparent band folding and peak-dip-hump structure. However, we argue that such scattering effects are difficult to reconcile with the observed change in the actual kink dispersion, as seen in the kink at the antinode. Increased scattering should also result in broader MDC widths but as seen in Fig.~\ref{figs:5}, the MDC width of cleave two is smaller than that of cleave one despite having weaker AF induced band folding.\\

\begin{figure*}[tbp]
\begin{center}
\includegraphics[scale=1.0]{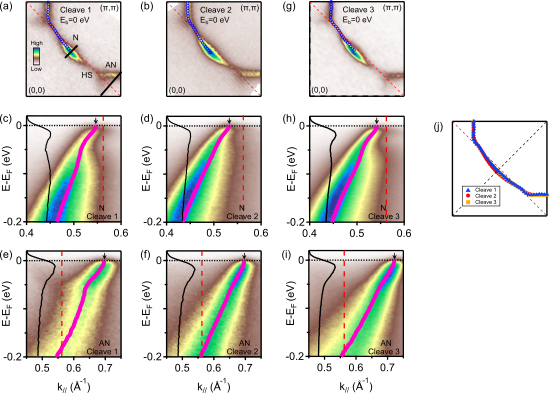}
\end{center}
\caption{\textbf{The interplay between electron-phonon coupling and AF fluctuation.} Panels (a) and (b) show the CES mapping at E$_\mathrm{F}$ for two different cleaves denoted 1 and 2. (c) and (d) illustrate the nodal energy-momentum cuts along the direction of N as indicated in panel (a) for cleaves 1 and 2, respectively. Similarly, (e) and (f) show the antinodal cuts along AN for the two cleaves. In each panel of (c)-(f), the $k_\mathrm{F}$-EDCs taken at the position marked by the black arrow are plotted as a black curve on the left edge of each panel. Panels (g)-(i) are additional plots with cleave 3 in addition to the Fig.~3 in the manuscript. Specifically, (g) The CES mapping at E$_\mathrm{F}$ for cleave 3. (h) and (i) illustrate the nodal and antinodal energy-momentum cuts along the direction of N and AN for cleave 3. (j) The comparison of the fitted \enquote{gossamer} FS for the three cleaves.}
\label{figs:6}
\end{figure*}

\clearpage

%%%%%%%%%%%%%%%%%% Part 4 %%%%%%%%%%%%%%%%%%%%%%%%%%%%%%%%%%%%%%%%%%%%%%

{\bf IV. Dispersion anomalies near the zone boundary}

\begin{figure*}[b]
\begin{center}
\includegraphics[scale=1.0]{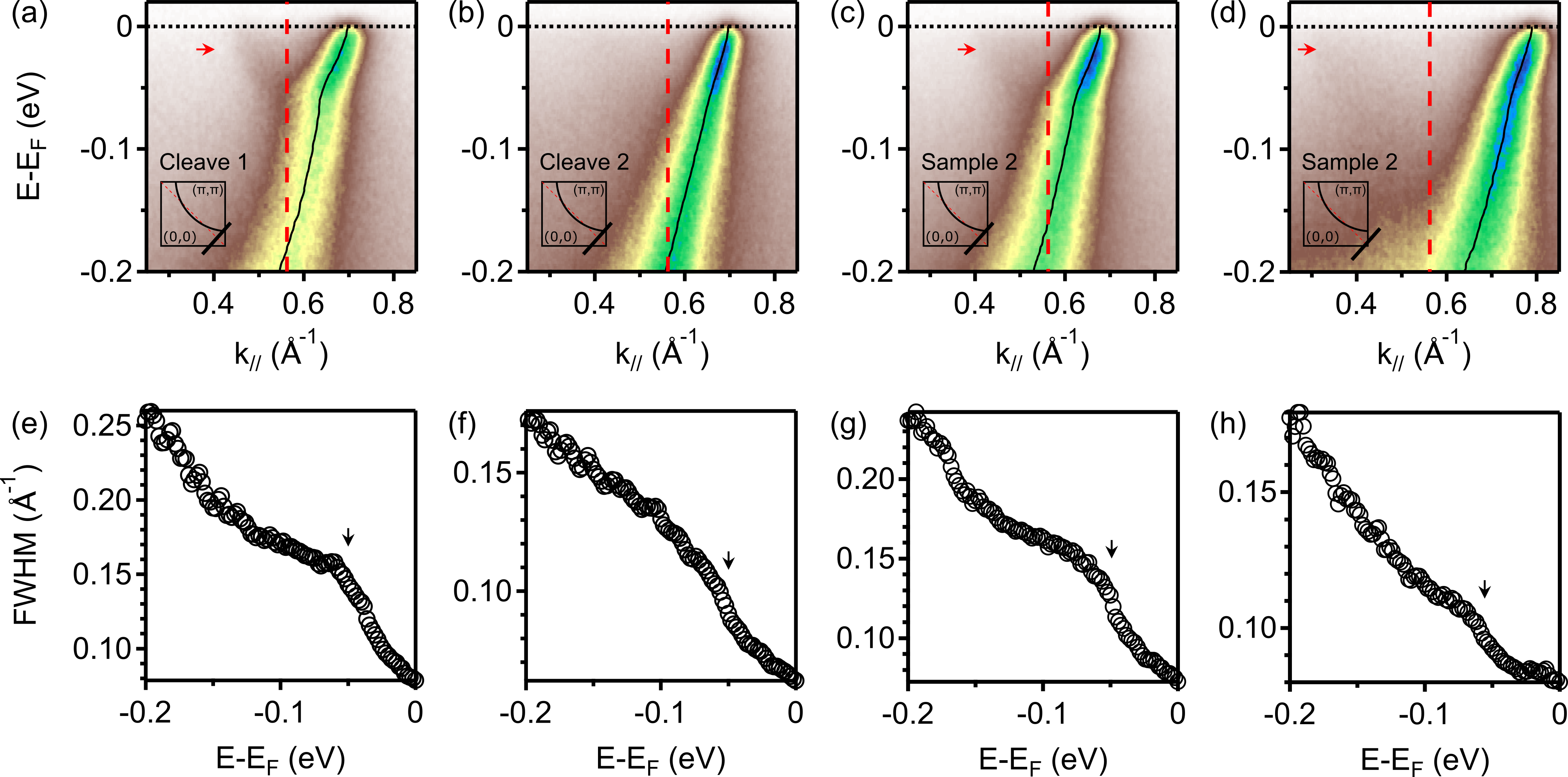}
\end{center}
\caption{\textbf{The kink near the zone boundary.} (a)-(d) The energy-momentum cuts are taken along the direction depicted in the inset, with details of the cleave and sample provided in the label. The red arrows points out the shadow band resulting from the AF folding. From each intensity plot, the full-width-at-half-maximum (FWHM) information is extracted from the MDC analysis and shown in the lower panels (e)-(h) respectively. The black arrows indicate the energy position of the kink.}
\label{figs:7}
\end{figure*}

In Fig.~\ref{figs:7}, we present energy momentum cuts near the zone boundary together with the information extracted from single-Lorentzian fits to MDCs in an energy range spanning [0, 0.2]~eV in binding energy ($E_\mathrm{b}$). The peak positions from the fits are represented by black curves in the intensity plot, while the width (FWHM) of the peaks is depicted in the lower panels. The results show traces of interaction occurring at $E_\mathrm{b} = 50\pm 5$~meV. Based on this analysis, we conclude that a prominent boson mode is observed near the zone boundary, with no other modes clearly resolved within this energy range. At first glance, the kink depicted in panel (a) might appear to be associated with the folding. However, we demonstrate in panels (c) and (d) for a different sample that the kink persists and remains unaffected even when the main dispersion is distant from the AFZB. In panels (e) and (g), an additional feature is observed at higher energies (around $E_\mathrm{b}=0.18$ eV), corresponding to the energy where the main band intersects the AFZB. This feature shows consistency with the visibility of the shadow band, marked by the red arrow in the upper panels. We consider this as further evidence of the intimate interplay between AF order and electron-phonon interaction.

%%%%%%%%%%%%%%%%%% Part 5 %%%%%%%%%%%%%%%%%%%%%%%%%%%%%%%%%%%%%%%%%%%%%%
{\bf V. The size of the AF gap}

\begin{figure*}[h]
\begin{center}
\includegraphics[scale=1.0]{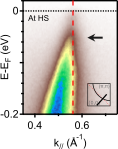}
\end{center}
\caption{\textbf{The energy-momentum cut around the hot spot (HS).} The arrow points to the AF feature.}
\label{figs:9}
\end{figure*}

The gap induced in the main band by AF fluctuations is most clearly observed at the hot-spot in the Brillouin zone, the point where the Fermi surface crosses the AF zone boundary. At this point the AF induced gap is symmetrical around the Fermi energy and the gap is most clearly observed. As seen from Fig.~\ref{figs:9} the gap measured from the Fermi energy is approximately 50 meV.

%%%%%%%%%%%%%%%%%% Part 6 %%%%%%%%%%%%%%%%%%%%%%%%%%%%%%%%%%%%%%%%%%%%%%

{\bf VI. The experimental spectral densities at the node and antinode}

\begin{figure*}[h]
\begin{center}
\includegraphics[scale=1.0]{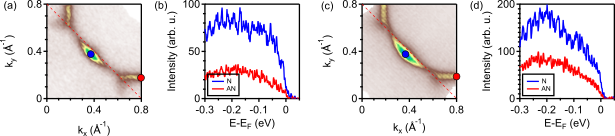}
\end{center}
\caption{\textbf{Experimental spectral weight at the node and antinode.} (a) Constant energy surface mapping at E$_\mathrm{F}$ for the cleave 1. (b) Momentum-integrated EDCs for cleave 1. The momentum positions for EDC extraction are indicated as circles in the inset of panel (a) with corresponding colors. The momentum integration is $0.2~\textup{\AA}^{\mathsf{-1}} \times 0.2~\textup{\AA}^{\mathsf{-1}}$. (c)--(d) presents the results for cleave 2.}
\label{figs:10}
\end{figure*}

The experimental spectral densities at the node and antinode for two different cleaves in NCCO are illustrated in Fig.~\ref{figs:10}. Panels (a) and (c) show constant energy surface maps at the Fermi energy (E$_\mathrm{F}$) for cleaves 1 and 2, respectively. Panels (b) and (d) display the momentum-integrated EDCs directly extracted from the mapping data for cleaves 1 and 2. Momentum integration was performed with a grid spacing of $0.2~\textup{\AA}^{-1} \times 0.2~\textup{\AA}^{-1}$ in the $\mathrm{k_x}$ and $\mathrm{k_y}$ directions to account for the broadening effect. The plotted EDCs have no vertical offset in each panel, indicating an increased density of states at the node.

\vspace{60mm}
\PRLsep
%\bibliography{ref}

\noindent\small[1]	\;M. Horio, T. Adachi, Y. Mori, A. Takahashi, T. Yoshida, H. Suzuki, L. Ambolode, K. Okazaki,\\ \indent\;\;\;\;\; K. Ono, H. Kumigashira, \textit{et al.}, Suppression of the antiferromagnetic pseudogap in the electron-\\ \indent\;\;\;\;\;doped high-temperature superconductor by protect annealing, Nature Communications, \textbf{7},\\ \indent\;\;\;\;\;10567 (2016).\\
\noindent\small[2]	\;N. Harima, J. Matsuno, A. Fujimori, Y. Onose, Y. Taguchi, and Y. Tokura, Chemical potential\\ \indent\;\;\;\;\; shift in $\mathrm{Nd_{1.85}Ce_{0.15}CuO_4}$: $\mathrm{C}$ontrasting behavior between the electron-and hole-doped cuprates,\\ \indent\;\;\;\;\;Physical Review B \textbf{64}, 220507 (2001).\\
\noindent\small[3]	\;K. Yamada, K. Kurahashi, T. Uefuji, M. Fujita, S. Park, S.-H. Lee, and Y. Endoh, Commen-\\ \indent\;\;\;\;\;surate spin dynamics in the superconducting state of an electron-doped cuprate superconductor,\\ \indent\;\;\;\;\;Physical Review Letters \textbf{90}, 137004 (2003).\\
\noindent\small[4]	\;G. Williams, J. Haase, M.-S. Park, K. H. Kim, and S.-I. Lee, Doping-dependent reduction of\\ \indent\;\;\;\; the Cu nuclear magnetic resonance intensity in the electron-doped superconductors, Physical\\ \indent\;\;\;\; Review B \textbf{72}, 212511 (2005).\\